\documentclass[prc,preprint]{revtex4}

\usepackage{graphicx}

\begin{document}

\title{\bf Maximum Mass of the Hot Neutron Star with the Quark Core}

\author{{\bf T. Yazdizadeh$^1$ and G. H. Bordbar$^{2,3}$}}
 \affiliation{
$^1$Islamic Azad University, Bafgh Branch, Bafgh, Iran\\
 $^2$Department of
Physics, Shiraz University,
Shiraz 71454, Iran\\
$^3$Research Institute for Astronomy and Astrophysics of Maragha,\\
P.O. Box 55134-441, Maragha, Iran }

%%%%%%%%%%%%%%%%%%%%%%%%%%%%%%%%%%%%%%%%%%%%%%%%%%%%%%%%%%%%%%%%%%%%%%%%%

%%%%%%%%%%%%%%%%%%%%%%%%%%%%%%%%%%%%%%%%%%%%%%%%%%%%%%%%%%%%%%%%%%%%

\begin{abstract}
We have considered a hot neutron star with a quark core, a mixed
phase of quark-hadron matter, and a hadronic matter crust and
have determined the equation of state of the hadronic phase and
the quark phase,  we have then found the equation of state of the
mixed phase under the Gibbs conditions. Finally, we have computed
the structure of hot neutron star with the quark core and compared
our results with those of the neutron star without the quark core.
For the quark matter calculations, we have used the MIT bag model
in which the total energy of the system is considered as the
kinetic energy of the particles plus a bag constant.
For the hadronic matter calculations, we have used the lowest
order constrained variational (LOCV) formalism.
Our calculations show that the results for the maximum
gravitational mass of the hot neutron star with the quark core are
substantially different from those of without the quark core.
\end{abstract}
%%%%%%%%%%%%%%%%%%%%%%%%%%%%%%%%%%%%%%%%%%%%%%%%%%%%%%%
%%%%%%%%%%%%%%%%%%%%%%%%%%%%%%%%%%%%%%%%%%%%%%%%%%%%%%%
\maketitle
%%%%%%%%%%%%%%%%%%%%%%%%%%%%%%%%%%%%%%%%%%%%%%%%%%%%%%%%%%%%%%%%%%%%
% \noindent {\bf Keywords:} hot neutron star, quark core, equation of
% state, structure, maximum mass, radius

\section{Introduction}
A hot neutron star is born following the gravitational collapse of
the core of a massive star just after the supernova explosion. The
interior temperature of a neutron star at its birth is of order
$20-50\ MeV$ \cite{burrows}. Therefore, the high temperature of
these stages cannot be neglected with respect to the Fermi
temperature throughout the calculation of its structure. This
shows that the equation of state of the hot dense matter is very
important for investigating the structure of a newborn neutron
star. Depending on the total number of nucleons, a newborn neutron
star evolves either to a black hole or to a stable neutron star
\cite{strobel}. Hence, calculation of the maximum mass of hot
neutron star is of special interest in astrophysics.

As we go from surface to the center of a neutron star, at
sufficiently high densities, the matter is expected to undergo a
transition from hadronic matter, where the quarks are confined
inside the hadrons, to a state of deconfined quarks. Finally,
there are up, down and strange quarks in the quark matter. Other
quarks have high masses and do not appear in this state.
Glendenning has shown that a proper construction of the
hadron-quark phase transition inside the neutron stars implies the
coexistence of nucleonic matter and quark matter over a finite
range of the pressure. Therefore, a mixed hadron-quark phase
exists in the neutron star and its energy is lower than those of
the quark matter and nucleonic matter \cite{glen1}. These show
that we can consider a neutron star as composed of a hadronic
matter layer, a mixed phase of quarks and hadrons and, in core, a
quark matter.
Recent Chandra observations also imply that the objects RX
J185635-3754 and 3C58 could be neutron stars with the quark core
\cite{prakash}.

Burgio et al. have investigated the structure of neutron stars
with the quark core at zero \cite{burgio1} and finite temperature
\cite{burgio2} with the Brueckner-Bethe-Goldstone formalism to
determine the equation of state of the hadronic matter, they have
used .
We have calculated the structure properties of the cold neutron
star by considering a quark phase at its core \cite{b1} and
compared the results with our previous calculations for the
neutron star without the quark core \cite{bh}. In these works, we
have employed the lowest order constrained variational (LOCV)
method for the hadronic matter calculations.
In the present paper, we intend to extend these calculations for
the hot neutron star with the quark core.
%%-----------------------------------------------------------
\section{Equation of State}

As it was mentioned in the previous section, we  consider a
neutron star composed of a hadronic matter (hadron phase), a mixed
phase of quarks and hadrons, and a quark core (quark phase).
Therefore, we should calculate the equation of state of these
phases separately as follows.

\subsection{Hadron Phase}
\label{HP}
 For this phase of the neutron star matter, we consider
the total energy per nucleon as the sum of contributions from the
leptons and nucleons,
\begin{equation}
E = E_{lep} + E_{nucl}\cdot
\end{equation}

The contribution from the energy of leptons (electrons and muons)
is
\begin{equation}
E_{lep} = E_e + E_\mu,
\end{equation}
where $E_e$ and $E_\mu$ are the energies of electrons and muons,
respectively,
\begin{equation}
E_{i} = \frac{m_i^4c^5}{\pi^2 n \hbar^3}\int_0^\infty
\frac{\sqrt{1 + x^2}}{ 1 + Exp\{\beta[m_i c^2\sqrt{1 + x^2} -
\mu_i ]\}} x^2 dx. \label{Elep}
\end{equation}
Here $\mu_i$ and $m_i$ are the chemical potential and mass of
particle $i$, $\beta = \frac{1}{k_B T}$ ($T$ is the temperature),
$n$ is the total number density of nucleons ($n = n_p + n_n$), $c$
is speed of light and $x$ is as follows,
\begin{equation}
x = \frac{\hbar k}{m_i c}.
\end{equation}

In our calculations, the equation of state of hot nucleonic matter
is determined using the lowest order constrained variational
(LOCV) method as follows
 \cite{b2, b3, b4, b5, b6, b7, b8, b9}. We adopt a trail
wave function as
\begin{equation}\label{eq7}
\psi=F\phi,
\end{equation}
where $\phi$ is the Slater determinant of the single-particle wave
function and $F$ is the correlation function which is taken to be
\begin{equation}\label{eq7}
F={\cal S}\prod_{i>j}f(ij).
\end{equation}
${\cal S}$ is a symmetrizing operator. For the energy of nucleonic
matter, we consider up to the two-body term in the cluster
expansion,
\begin{equation}\label{eq7}
E_{nucl}=E_1+E_2.
\end{equation}

The one body term $E_1$ for the hot asymmetrical nucleonic matter
that consists of $Z$ protons and $N$ neutrons is simply the fermi
gas kinetic energy,
 \begin{equation}\label{eq7}
E_1=\sum_{i=1,2}{\cal E}_i
\end{equation}
Label 1 and 2 are used instead of proton and neutron,
respectively, and ${\cal E}_i$ is
\begin{equation}\label{eq7}
{\cal E}_i=\sum_k\frac{\hbar^2k^2}{2m_i}f_i(k, T,n_i),
\end{equation}
where $f(k, T, n_i)$ is the Fermi-Dirac distribution function
\cite{fetter},
\begin{equation}\label{eq7}
f(k, T,n_i)=\frac{1}{e^{\beta[\epsilon_i(k,T,n_i)-\mu_i( T,
n_i)]}+1}. \label{FD}
\end{equation}
In the above equation, $n_i$ are the number densitis and
$\epsilon_i$ are the single particle energies associated with the
protons and neutrons,
\begin{equation}\label{eq7}
\epsilon_i(k, T,n_i)=\frac{\hbar^2k^2}{2m_i^\ast( T,n_i)},
\end{equation}
where $m_i^\ast$ are the effective masses.

The two-body energy, $E_2$, is
\begin{equation}\label{eq7}
E_2=\frac{1}{2A}\sum_{ij}<ij|\nu(12)|ij-ji>,
\end{equation}
where
\begin{equation}\label{eq7}
\nu(12)=-\frac{\hbar^2}{2m}[f(12),[\nabla^2_{12},f(12)]]+f(12)V(12)f(12).
\end{equation}
$f(12)$ and $V(12)$ are the two-body correlation and
inter-nucleonic potential.

We note that the conditions of charge neutrality and beta
stability impose the following constraints on the number densities
and chemical potentials,
\begin{eqnarray}
n_p = n_e + n_\mu\\
\mu_n - \mu_p = \mu_e = \mu_\mu.
\end{eqnarray}
 The procedure to calculate the nucleonic matter has been fully discussed in the Refs.
\cite{b2,b3}.

\subsection{Quark Phase}\label{QP}
We use the MIT bag model for the quark matter calculations. In
this model, the energy density is the  kinetic energy of quarks
plus a bag constant (${\cal B}$) which is interpreted as the
difference between the energy densities of non interacting quarks
and interacting ones \cite{farhi1},

\begin{equation}\label{eq6}
    {\cal E}_{tot} = {\cal E}_u + {\cal E}_d + {\cal E}_s + {\cal B},
\end{equation}
where ${\cal E}_i$ is the kinetic energy per volume of particle
$i$,
\begin{equation}\label{eq7}
{\cal
E}_i=\frac{g}{2\pi^2}\int_0^{\infty}{(m_i^2c^4+\hbar^2k^2c^2)^{1/2}}
{f(k, T, n_i)}{k^2dk}.
\end{equation}
In above equation, $g$ is the degeneracy number of the system and
$n_i$ is the number density of particle $i$,
\begin{equation}\label{eq7}
n_i=\frac{g}{2\pi^2}\int_0^{\infty} {f(k, T, n_i)}{k^2dk}.
\end{equation}
For the quark phase, the Fermi-Dirac distribution function, $f(k,
T, n_i)$, is given by
\begin{equation}\label{fermi}
f(k, T,
n_i)=\frac{1}{Exp\{\beta((m_i^2c^4+\hbar^2k^2c^2)^{1/2}-\mu_i)\}+1}.
\end{equation}
We assume that the up and down quarks are massless, the strange
quark has a mass equal to $150\ MeV$ and ${\cal B}=90\ MeV
fm^{-3}$.
Now, by applying the beta stability and charge neutrality
conditions, we get the following relations for the chemical
potentials and number densities,
\begin{equation}\label{eq7}
\mu_d=\mu_u+\mu_l,
\end{equation}
\begin{equation}\label{eq6}
\mu_s=\mu_u+\mu_l,
\end{equation}
\begin{equation}\label{eq6}
\Rightarrow\mu_d=\mu_s,
\end{equation}
\begin{equation}\label{eq6}
\frac{2}{3}n_u-\frac{1}{3}n_d-\frac{1}{3}n_s-n_l=0,
\end{equation}
\begin{equation}\label{eq6}
n_B=\frac{1}{3}(n_u+n_d+n_s),
\end{equation}
where $n_l$ and $\mu_l$ are the leptonic number density and
chemical potential, and $n_B$ is the baryonic number density.

The pressure of the system is calculated from free energy using
the following equation,
\begin{equation}\label{eq11}
P=\sum_i {n_i \frac{\partial {\cal F}_i}{\partial n_i}-{\cal
F}_i},
\end{equation}
where the Helmholtz free energy per volume (${\cal F}$) is given
by
\begin{equation}\label{eq10}
    {\cal F} = {\cal E}_{tot} - T{\cal S}_{tot}.
\end{equation}
The entropy of quark matter (${\cal S}_{tot}$) can be written as
follows,
\begin{equation}\label{eq6}
    {\cal S}_{tot}={\cal S}_u + {\cal S}_d + {\cal S}_s,
\end{equation}
where ${\cal S}_i$ is the entropy of particle $i$,
\begin{eqnarray}\label{eq8}
{\cal S}_i(n_i, T)&=&-\frac{3}{\pi^2}k_B\int_0^{\infty} [f(k, T,
n_i)\ln(f(k, T, n_i))
    \nonumber\\&&
    +(1-f(k, T, n_i))\ln(1-f(k, T, n_i))]k^2dk.
\end{eqnarray}

\subsection{Mixed phase}\label{MP}
For the mixed phase, where the fraction of space occupied by quark
matter smoothly increases from zero to unity, we have a mixture of
hadrons, quarks and electrons. In the mixed phase, according the
Gibss equilibrium condition, the temperatures, pressures and
chemical potentials of the hadron phase (H) and quark phase (Q)
are equal \cite{glen1}. Here, for each temperature we let the
pressure to be an independent variable.

The Gibss conditions implies that
\begin{equation}\label{eq6}
\mu^Q_n=\mu^H_n,
\end{equation}
\begin{equation}\label{eq6}
\mu^Q_p=\mu^H_p,
\end{equation}
where $\mu^H_n$ and $\mu^Q_n$ ($\mu^H_p$ and $\mu^Q_p$) are the
neutron (proton) chemical potentials in the hadron phase and the
quark phase, respectively,
\begin{equation}\label{eq30}
\mu_n=\frac{\partial{\cal E}}{\partial{n_n}},
\end{equation}
\begin{equation}\label{eq31}
\mu_p=\frac{\partial{\cal E}}{\partial{n_p}}.
\end{equation}
In above equations, ${\cal E}$ is the energy density of the
system,
\begin{equation}\label{eq32}
{\cal E}=n(E+mc^2).
\end{equation}

To obtain $\mu^H_p$ and $\mu^H_n$ for the hadronic matter in
mixed phase, we use the semiempirical mass formula
\cite{kutschera,langris,wiringa},
\begin{equation}\label{eq33}
E=T(n,x)+V_0(n)+(1-2x)^2V_2(n),
\end{equation}
where $x=\frac{n_p}{n}$ is the proton fraction. $T(n,x)$ is
kinetic energy contribution and the functions $V_0$ and $V_2$
represent the interaction energy contributions which are
determined  from  the energies of the symmetric nuclear matter
$(x=\frac{1}{2})$ and pure neutron matter $(x = 0)$.
We calculate $V_0$ and $V_2$ using our results for the LOCV
calculation of nucleonic matter with the $UV_{14} + TNI$ nuclear
potential which is discussed in section \ref{HP}. Now, we can
obtain the chemical potentials of neutrons and protons from Eqs.
(\ref{eq30})-(\ref{eq33}) as follows,
\begin{eqnarray}\label{eq34}
\mu^H_p&=&T(n,x)+n\frac{\partial{T(n,x)}}{\partial{n}}+\frac{\partial{T(n,x)}}
{\partial{x}}+V_0(n)+n{V_0}^\prime(n)\nonumber\\
&&+(-3+8x-4x^2)V_2(n)+ (1-2x)^2n{V_2}^\prime(n)+mc^2,
\end{eqnarray}

\begin{eqnarray}\label{eq34}
\mu^H_n&=&T(n,x)+n\frac{\partial{T(n,x)}}{\partial{n}}-\frac{\partial{T(n,x)}}
{\partial{x}}+V_0(n)+n{V_0}^\prime(n)\nonumber\\&&+(1-4x^2)V_2(n)+
(1-2x)^2n{V_2}^\prime(n)+mc^2.
\end{eqnarray}

For the quark matter in mixed phase, we have
\begin{equation}\label{eq35}
\mu_p^Q=2\mu_u+\mu_d,
\end{equation}
\begin{equation}\label{eq36}
\mu_n^Q=\mu_u+2\mu_d.
\end{equation}
At a certain pressure, we calculate $\mu_u$ for different $\mu_d$
under the condition that the densities yield this certain
pressure. By calculating  $\mu_u$ and $\mu_d$, we obtain $\mu^Q_p$
and $\mu^Q_n$.

Now, we plot $\mu_p$ versus $\mu_n$ for both hadron and quark
phases, the cross point of the two curves satisfies the Gibss
conditions. In the mixed phase, as the chemical potentials
determine the densities, the volume fraction occupied by quark
matter, $\chi $, can be obtained by the requirement of global
charge neutrality,
\begin{equation}\label{eq6}
\chi(\frac{2}{3}n_u-\frac{1}{3}n_d-\frac{1}{3}n_s)+(1-\chi)n_p-n_e=0.
\end{equation}
Finally, we can calculate the baryonic density of the mixed phase
(M),
\begin{equation}\label{eq6}
n_B=\chi n_Q+(1-\chi)n_H,
\end{equation}
and then the total energy density of mixed phase is found,
\begin{equation}\label{eq6}
{\cal E}_{M}=\chi{\cal E}_{Q}+(1-\chi){\cal E}_{H}.
\end{equation}

\subsection{Results}

\begin{figure}
\includegraphics{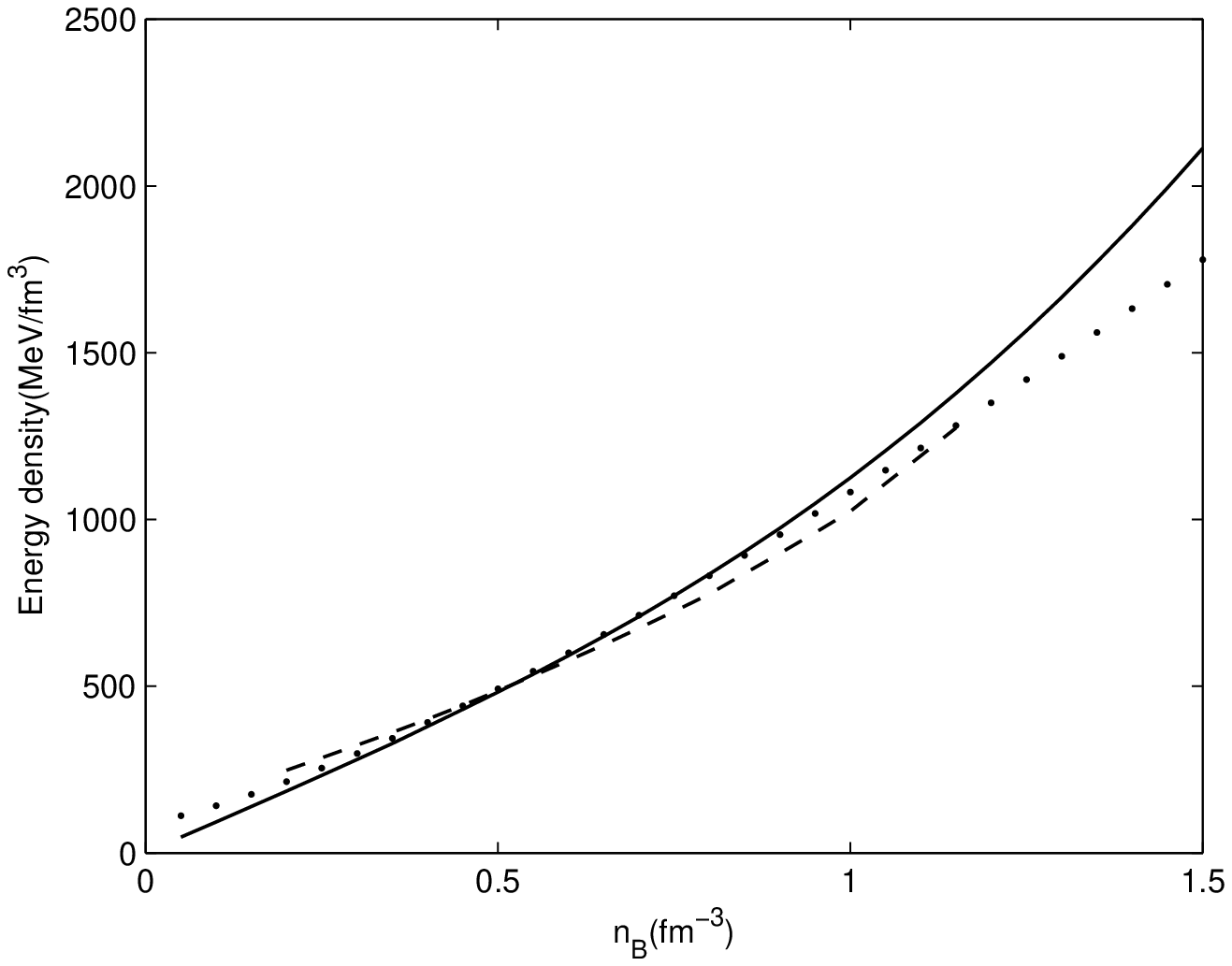}
 \caption{
Energy density versus the baryonic density  at $T=10\ MeV$} for
the hadron phase (solid curve), quark phase (dotted curve) and
mixed phase (dashed curve). \label{en1}
\end{figure}

\begin{figure}
\includegraphics{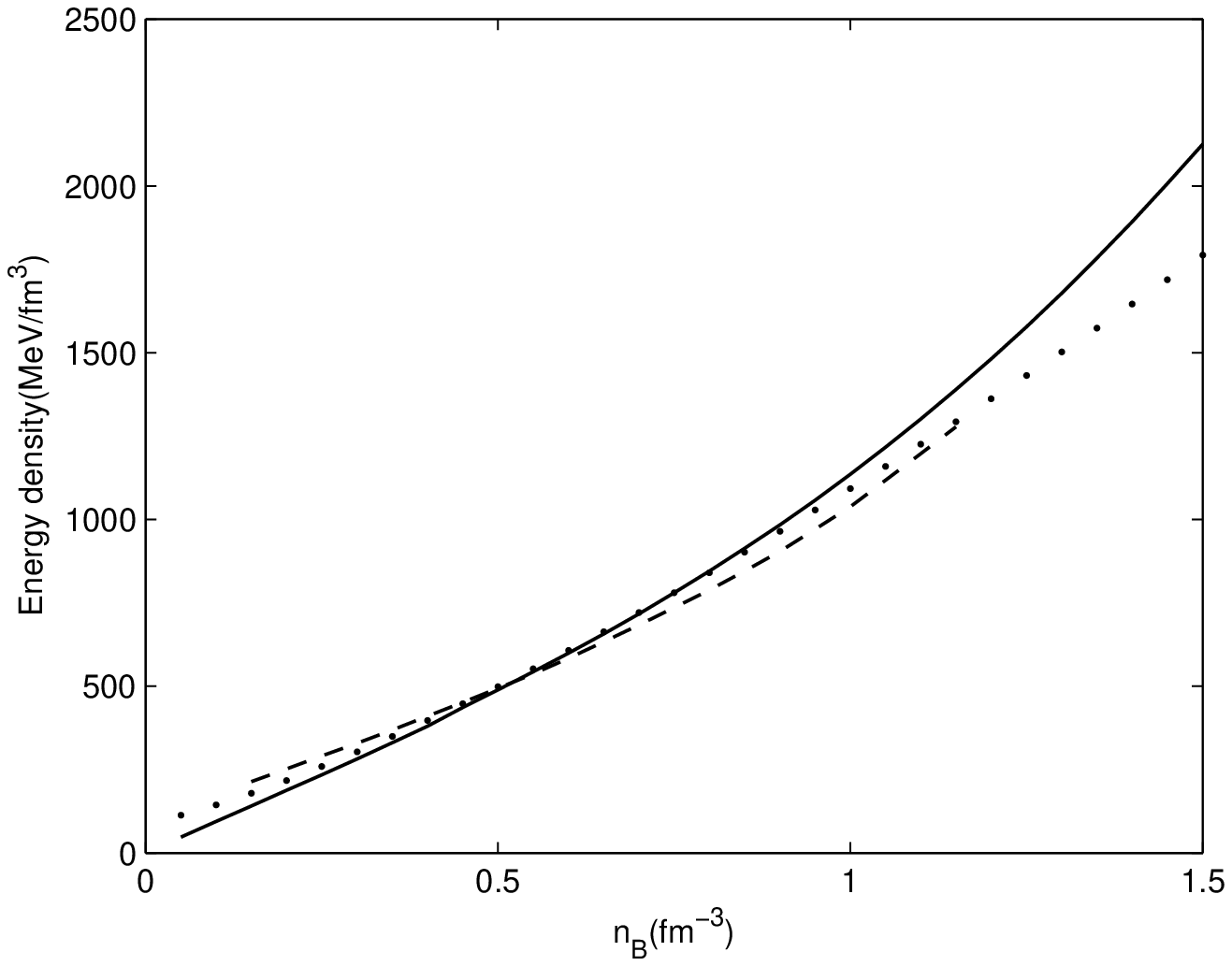}
 \caption{As
Fig. \ref{en1} but at $T=20\ MeV$.} \label{en2}
\end{figure}

We have shown our results for the energy densities of hadron
phase, quark phase and mixed phase in Figs. \ref{en1} and
\ref{en2} at two different temperatures.
Figs. \ref{en1} and \ref{en2} show that at low densities the
energy density of the hadronic matter is lower than those of other
phases. However, as the density increases, at first the energy of
mixed phase and finally the energy of quark phase is  lower than
those of other phases.
We also see that there is a mixed phase for a range of densities.
Below (beyond) this range, we have the pure hadron (quark) phase.
By comparing Figs. \ref{en1} and \ref{en2}, we see that for a
given value of the density, the energies of all phases increases
by increasing the temperature.

Using the above calculated energy density, we can determine the
equation of state and finally the structure of the hot neutron
star with the quark core which is discussed in the next section.

\section{Structure of the Hot Neutron Star with the Quark Core }
The structure of neutron star is determined by numerically
integrating the Tolman-Oppenheimer-Volkoff equation (TOV)
\cite{shapiro,glen2,weber,alder},

\begin{equation}\label{eq12}
    \frac{dP}{dr}=-\frac{G[{\cal E}(r)+\frac{P(r)}{c^2}]
    [m(r)+\frac{4\pi r^3 P(r)}{c^2}]}{r^2
    [1-\frac{2Gm(r)}{rc^2}]},
\end{equation}

\begin{equation}\label{eq13}
    \frac{dm}{dr}=4\pi r^2{\cal E}(r),
\end{equation}
where $P$ is the pressure and ${\cal E}$ is the total energy
density. For a given equation of state in the form $P({\cal E})$,
the TOV equation yields the mass and radius of star as a function
of the central mass density.

In our calculations for the structure of hot neutron star with the
quark core, we use the following equations of state: (i) Below the
density of $0.05\ fm^3$, we use the equation of state calculated
by Baym \cite{baym}. (ii) From the density of $0.05\ fm^3$ up to
the density where the mixed phase starts, we use the equation of
state of pure hadron phase calculated in section \ref{HP}. (iii)
In the range of densities in which there is the mixed phase, we
use the equation of state calculated in section \ref{MP}. (iv)
Beyond the density of end point of the mixed phase, we use the
equation of state of pure quark phase calculated in section
\ref{QP}.
All calculations are done for ${\cal B}=90\ MeV fm^{-3}$ at two
different temperatures $T=10$ and $20\ MeV$. Our results are as
follows.

\begin{figure}
\includegraphics{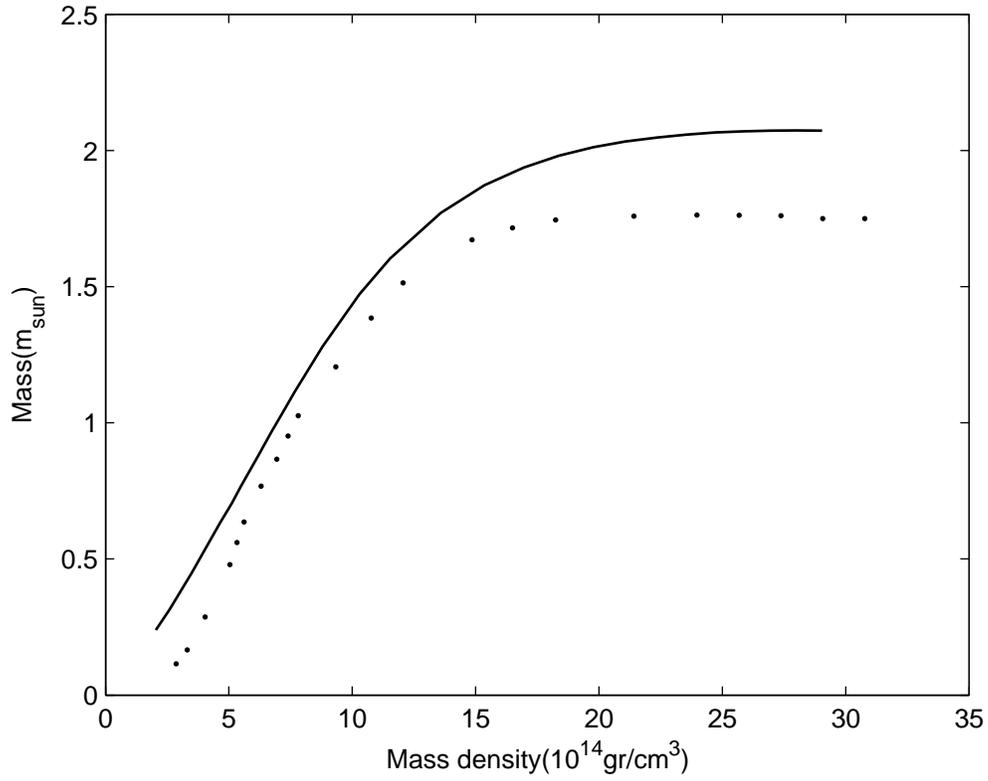}

\caption{Gravitational mass versus the central mass density for
the neutron star with (dotted curve) and neutron star without
 (solid curve) the quark core at $T=10\ MeV$.} \label{me1}
\end{figure}

\begin{figure}
\includegraphics{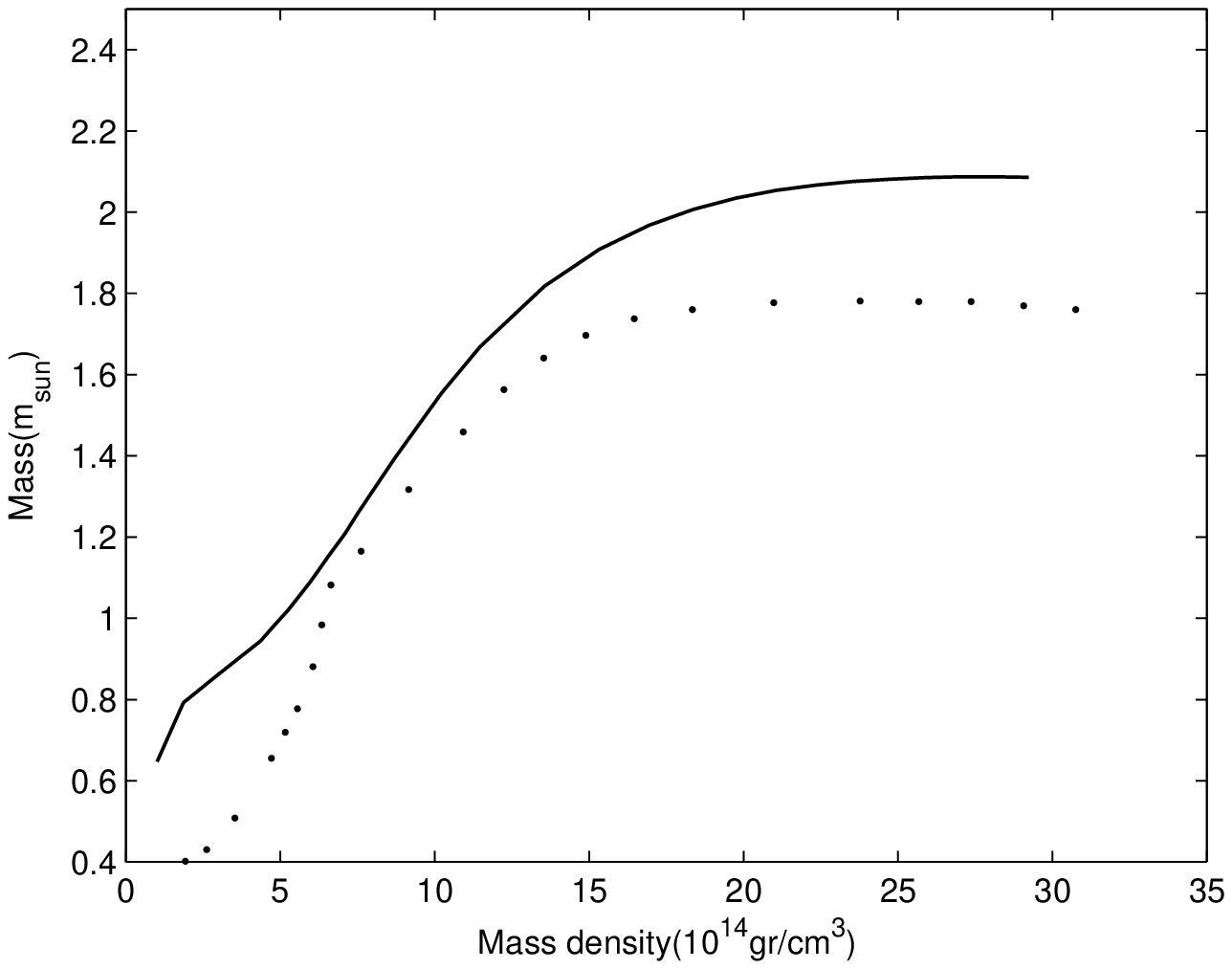}
\caption{As Fig. \ref{me1} but at $T=20\ MeV$.} \label{me2}
\end{figure}

\begin{figure}
\includegraphics{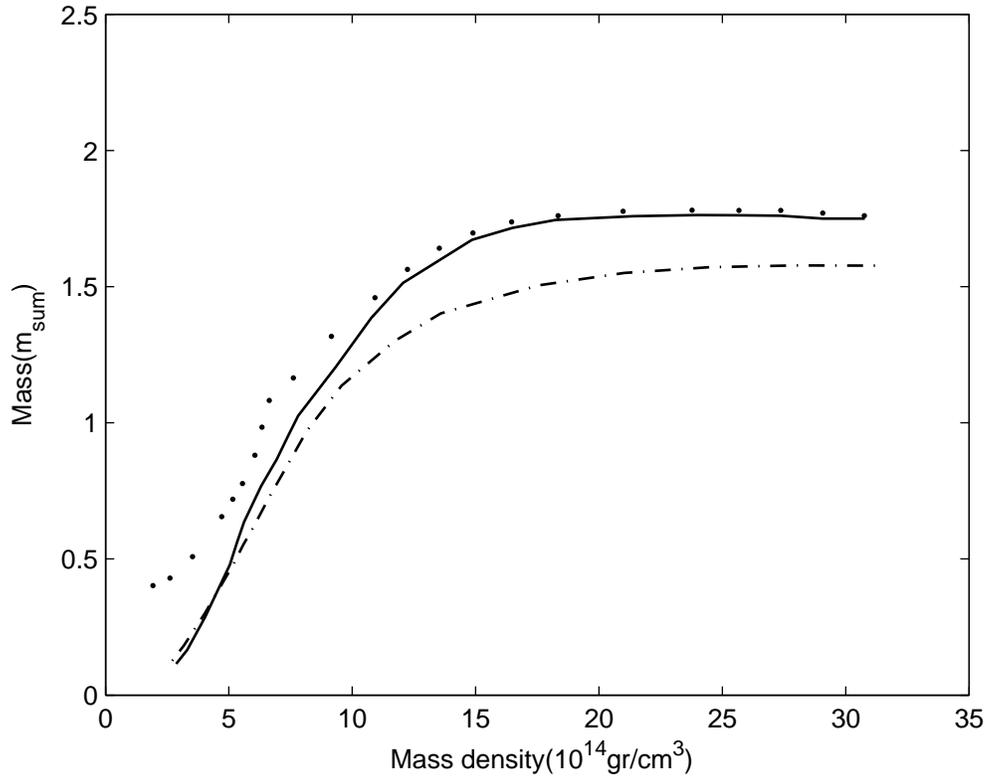}
\caption{Gravitational mass versus the central mass density for
the neutron star with the quark core at $T=0$ (dotted-dashed
curve), $10$ (solid curve) and $20\ MeV$ (dotted curve).}
\label{me120}
\end{figure}

The gravitational mass as a function of the central mass density
for the hot neutron star with the quark core  at two different
temperatures has been presented in Figs. \ref{me1} and \ref{me2}.
It is seen that for both relevant temperatures, the gravitational
mass increases by increasing the central mass density and finally
reaches a limiting value (maximum mass).
In Figs. \ref{me1} and \ref{me2}, our results for the case of
neutron star without the quark core have been also given for
comparison. We see that by including the quark core for the
neutron star, our results for the gravitational mass are
substantially affected.
For the neutron star with the quark core, our results for the
gravitational mass at three different temperatures ($T=0$, $10$
and $20\ MeV$) have been compared in Fig. \ref{me120}. It is seen
that the gravitational mass increases by increasing the
temperature.

\begin{figure}
\includegraphics{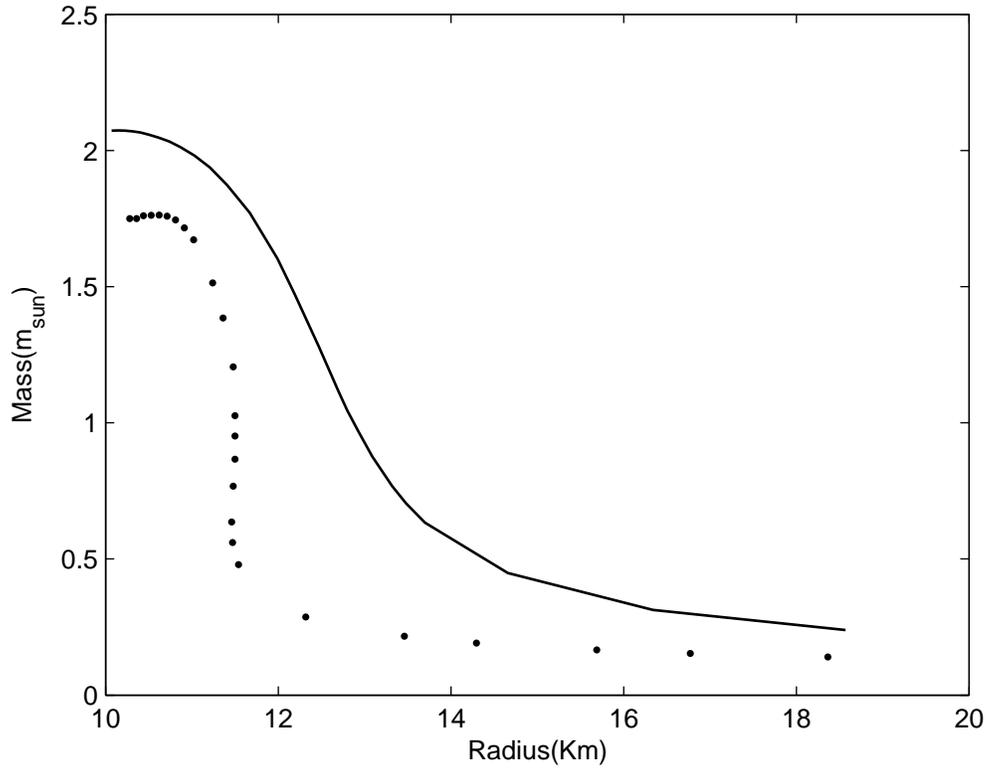}
\caption{ Mass-radius relation for the neutron star with (dotted
curve) and without (solid curve) the quark core at $T=10\ MeV$.}
\label{mr1}
\end{figure}

\begin{figure}
\includegraphics{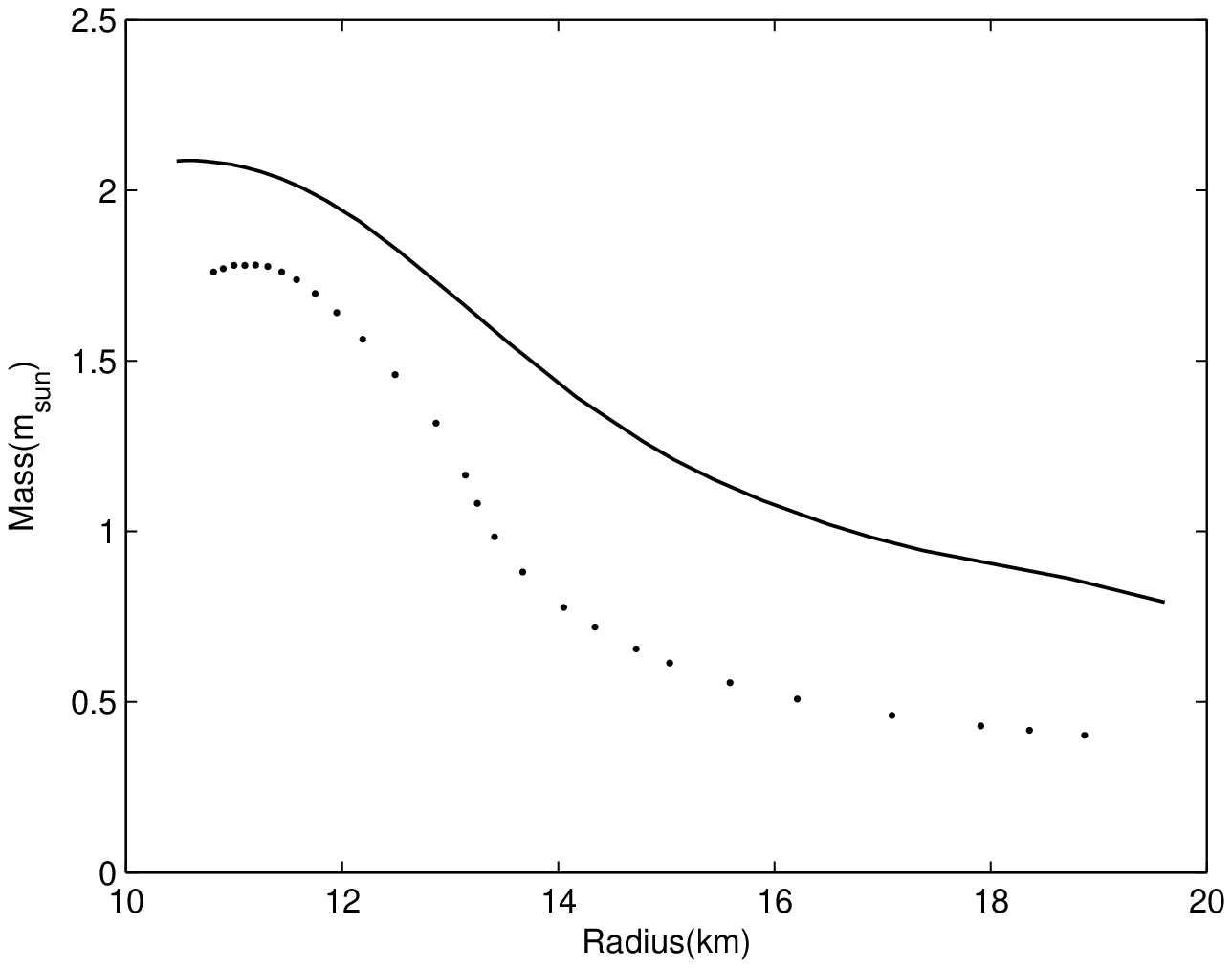}
\caption{As Fig. \ref{mr1} but at $T=20\ MeV$.} \label{mr2}
\end{figure}

Figs. \ref{mr1} and \ref{mr2} show the gravitational mass versus
the radius for both cases of the neutron star with and without the
quark core at two different temperatures. At each temperature, it
is seen that there is a reasonable difference between the
mass-radius relations of these two cases of the neutron star.
However, for both cases, we see that the radius decreases as the
mass increases. By comparing Figs. \ref{mr1} and \ref{mr2}, we can
see that the decreasing rate of the radius versus the mass is
substantially different for different temperatures.

Our results for the maximum gravitational mass of the hot neutron
star with the quark core and the corresponding values of radius
and central mass density have been given in Tables \ref{1} and
\ref{2} for two different temperatures. Our results for the case
of  hot neutron star without quark core have been also presented
for comparison. For different temperatures, it is seen that the
inclusion of the quark core considerably reduces the maximum mass
of the hot neutron star. This is due to the fact that by including
the quark core for the neutron star, the equation of state becomes
softer than that without the quark core. However, we do not see
any substantial changes for the radius and central mass density of
these two cases of the hot neutron star.

\section{Summary and Conclusion}
For the hot neutron star, from the surface toward the center, we
have considered a pure hadronic matter layer, a mixed phase of
quarks and hadrons in a  range of densities which are determined
by employing the Gibbs conditions, and a pure quark matter in the
core, to calculate its equation of state at finite temperature.
For calculating the equation of state of the hot hadronic matter,
we have applied the lowest order constrained variational (LOCV)
method at finite temperature. The equation of state of the hot
quark matter has been computed using the MIT bag model with the
bag constant ${\cal B}=90\ MeV fm^{-3}$.
Using this equation of state, we have solved the TOV equation by
numerical method to determine the structure properties of the hot
neutron star with the quark core at $T=10$ and $20\ MeV$. Then, we
have compared the results of these calculations with those for the
hot neutron star without the quark core.
It is found that our results for the maximum gravitational mass of
the neutron star with a quark core are less than those of the
neutron star without the quark core.

\acknowledgements{This paper is derived from research project
entitled: Structure of neutron stars with a quark core at finite
temperature. Financial support from Islamic Azad University, Bafgh
Branch, Research council is gratefully acknowledged. G. H. Bordbar
wishes to thank the Shiraz University Research council. G. H.
Bordbar also wishes to thank the Research Institute for Astronomy
and Astrophysics of Maragha for its financial support.}
%%%%%%%%%%%%%%%%%%%%%%%%%%%%%%%%%%%%%%%%%%%%%%%%%%%%%%%%%%%%%%%%%%%%
\newpage

%%%%%%%%%%%%%%%%%%%%%%%%%%%%%%%%%%%%%%%%%%%%%%%%%%%%%%%%%%%%%%%%%%%%

%%%%%%%%%%%%%%%%%%%%%%%%%%%%%%%%%%%%%%%%%%%%%%%%%%%%%%%%%%%%%%%%%%%%%%%%%%%%%%%%%%%%%%%%%%%%%%%%%%%%%%%
\newpage
\begin{table}
  \centering
  \caption{Maximum gravitational mass $(M_{\max })$, and the corresponding radius ($R$)
and central mass density  $(\varepsilon_{c})$ of the hot neutron
star without (NS) and with (NS+Q ) the quark core at $T=10\
MeV$.}\label{1}
  \begin{tabular}{cccc}
  \hline
   & $M_{\max }(M_{\odot})$ & $R(Km)$ & $\varepsilon_{c}(10^{14}gr/cm^3)$ \\
  \hline
  NS & 2.07 & 10.22 & 26.94 \\
  NS+Q& 1.76 & 10.45 & 27.38 \\
  \hline
\end{tabular}
\end{table}
%%%%%%%%%%%%%%%%%%%%%%%%%%%%%%%%%%%%%%%%%%%%%%%%%%%%%%%%%%%%%%%%%%%%%%%%%%%%%%%%%%%%%
\newpage
\begin{table}
  \centering
  \caption{As Table \ref{1} but  at $T=20\ MeV$.}\label{2}
  \begin{tabular}{cccc}
  \hline
   & $M_{\max }(M_{\odot})$ & R(Km) & $\varepsilon_{c}(10^{14}gr/cm^3)$ \\
  \hline
  NS & 2.09 & 10.64 & 27.01 \\
  NS+Q& 1.78 & 11 & 27.37 \\
  \hline
\end{tabular}
\end{table}
%%%%%%%%%%%%%%%%%%%%%%%%%%%%%%%%%%%%%%%%%%%%%%%%%%%%%%%%%%%%%%%%%%%%%%%%%%%%%%%%%%%%%
\newpage
%%%%%%%%%%%%%%%%%%%%%%%%%%%%%%%%%%%%%%%%%%%%%%%%%%%%%%%%%%%%%%%%%%%%%%%%%%%%%%%%%%%%%
%%%%%%%%%%%%%%%%%%%%%%%%%%%%%%%%%%%%%%%%%%%%%%%%%%%%%%%%%%%%%%%%%%%%%%%%%%%%%%%%%%%%%
\newpage
%%%%%%%%%%%%%%%%%%%%%%%%%%%%%%%%%%%%%%%%%%%%%%%%%%%%%%%%%%%%%%%%%%%%%%%%%%%%%%%%%%%%%
%%%%%%%%%%%%%%%%%%%%%%%%%%%%%%%%%%%%%%%%%%%%%%%%%%%%%%%%%%%%%%%%%%%%%%%%%%%%%%%%%%%%%
%%%%%%%%%%%%%%%%%%%%%%%%%%%%%%%%%%%%%%%%%%%%%%%%%%%%%%%%%%%%%%%%%%%%%%%%%%%%%%%%%%%%%
%%%%%%%%%%%%%%%%%%%%%%%%%%%%%%%%%%%%%%%%%%%%%%%%%%%%%%%%%%%%%%%%%%%%%%%%%%%%%%%%%%%%%
%%%%%%%%%%%%%%%%%%%%%%%%%%%%%%%%%%%%%%%%%%%%%%%%%%%%%%%%%%%%%%%%%%%%%%%%%%%%%%%%%%%%%
%%%%%%%%%%%%%%%%%%%%%%%%%%%%%%%%%%%%%%%%%%%%%%%%%%%%%%%%%%%%%%%%%%%%%%%%%%%%%%%%
\end{document}